\documentclass{aa}
\usepackage{graphicx}
\usepackage{hyperref}
\usepackage{txfonts}
\usepackage{xcolor}
\graphicspath{{Figures/}}

\begin{document} 

\title{Maximal growth rate of the ascending phase of a sunspot cycle for predicting its amplitude}
\titlerunning{Maximal growth rate for predicting solar cycle amplitude}
\authorrunning{T. Podladchikova et al.}

   \author{Tatiana Podladchikova\inst{1}
   \and Shantanu Jain\inst{1}
   \and Astrid M. Veronig\inst{2,3}
   \and Olga Sutyrina\inst{1}
   \and Mateja Dumbovi\'c\inst{4}
   \and Fr\'ed\'eric Clette\inst{5}
   \and Werner P\"otzi\inst{3}
      }

\institute{Skolkovo Institute of Science and Technology, Bolshoy Boulevard 30, bld. 1, 121205, Moscow, Russia \\
\email{t.podladchikova@skoltech.ru}
\and
University of Graz, Institute of Physics, Universit\"atsplatz 5, 8010 Graz, Austria\
\and
University of Graz, Kanzelh\"ohe Observatory for Solar and Environmental Research, Kanzelh\"ohe 19, 9521 Treffen, Austria
\and
Hvar Observatory, Faculty of Geodesy, University of Zagreb, Kaciceva 26, HR-10000, Zagreb, Croatia
\and
World Data Center SILSO, Royal Observatory of Belgium, Avenue Circulaire 3, 1180, Brussels, Belgium
}

\date{Received Month, Year; accepted Month, year}

\abstract
{Forecasting the solar cycle amplitude is important for a better understanding of the solar dynamo as well as for many space weather applications. Different empirical relations of solar cycle parameters with the peak amplitude of the upcoming solar cycle have been established and used for solar cycle forecasts, as, for instance, the Waldmeier rule relating the cycle rise time with its amplitude, the polar fields at previous minimum, and so on. Recently, a separate consideration of the evolution of the two hemispheres revealed even tighter relations.}
{We aim to introduce the maximal growth rate of sunspot activity in the ascending phase of a cycle as a new and reliable precursor of a subsequent solar cycle amplitude. We also intend to investigate whether the suggested precursor provides benefits for the prediction of the solar cycle amplitude when using the sunspot indices (sunspot numbers, sunspot areas) derived separately for the two hemispheres compared to the total sunspot indices describing the entire solar disc.}
{We investigated the relationship between the maximal growth rate of sunspot activity in the ascending phase of a cycle and the subsequent cycle amplitude on the basis of four data sets of solar activity indices: total sunspot numbers, hemispheric sunspot numbers from the new catalogue from 1874 onwards, total sunspot areas, and hemispheric sunspot areas.} 
{For all the data sets, a linear regression based on the maximal growth rate precursor shows a significant correlation. Validation of predictions for cycles 1-24 shows high correlations between
the true and predicted cycle amplitudes reaching $r=0.93$ for the total sunspot numbers. The lead time of the predictions varies from 2 to 49 months, with a mean value of 21 months. Furthermore, we demonstrated that the sum of maximal growth rate indicators determined separately for the north and the south hemispheric sunspot numbers provides more accurate predictions than that using total sunspot numbers. The advantages reach 27\% and 11\% on average in terms of rms and correlation coefficient, respectively. The superior performance is also confirmed with hemispheric sunspot areas with respect to total sunspot areas.}
{The maximal growth rate of sunspot activity in the ascending phase of a solar cycle serves as a reliable precursor of the subsequent cycle amplitude. Furthermore, our findings  provide a strong foundation for supporting regular monitoring, recording, and predictions of solar activity with hemispheric sunspot data, which capture the asymmetric behaviour of the solar activity and solar magnetic field and enhance solar cycle prediction methods.}


\maketitle

\section{Introduction}
The Sun is our nearest star and continuously provides our planet with energy, light, and heat, thus making it a very habitable environment for life. However, it is also the source of powerful explosions, which can affect modern technologies in space and on Earth. The Sun’s magnetic field drives the 11-year solar cycle and predicting its amplitude is of scientific and practical importance for many space weather applications. Most of the solar cycle prediction methods, which are extensively discussed in the  recent review by \citet{Petrovay2020}, are based on the sunspot number, as this index provides the longest standardised record of solar activity, spanning 300 years. The sunspot number series is maintained and continuously extended by the World Data Centre Sunspot Index and Long-term Solar Observations (WDC SILSO). Sunspot numbers are used for predictions on different timescales including short-, medium-, and long-term lead times \citep[see definitions in][]{Petrova2021}. Long-term predictions mainly concentrate on forecasting the amplitude of the next cycle by using a specific sunspot number precursor from a previous cycle, which, as pointed out by \citet{Petrovay2020}, and can also provide physical insights \citep{Ramaswamy1977, LantosSkewness2006, Kane2008, PodladchikovaLefebvreLinden2008, Podladchikova2011, Podladchikova2017, Brajsa2022}. Other precursors in use are based on geomagnetic activity \citep{OhlOhl1979, Feynman1982, GonzalezSchatten1988, Thompson1993, WilsonHathawayReichmann1998, Miao2020, Burud2021} and the Sun's polar fields \citep{SchattenDynamo1978, Schatten1996, Schatten2005, Svalgaard2005, WangSheeley2009, MunozJaramillo2012,Pesnell2018,Bisoi2020}. Several authors presented statistical methods for the prediction of the solar cycle amplitude \citep{Macpherson1995, Fessant1996, Zhang1996, Conway1998, Sello2001, Aguirre2008, Braja2009, Liu2010, Attia2013, Kitiashvili2016, Singh2017, Sarp2018, Kakad2020}. Over the last two decades, flux transport and dynamo models were also developed into physics-based prediction models \citep{Nandy2002,DikpatiGilman2006,CameronSchlussler2007,Choudhuri2007,Henney2012,CameronSchlussler2015,Jiang2018,Labonville2019}. Recently, \citet{Bhowmik2018} presented the first data-driven simulations of solar activity that extend the prediction window to a decade. 

Methods of short- and medium-term predictions with lead times from days to months mainly provide continuous forecast of an ongoing solar cycle. Currently, SILSO delivers the operational medium-term prediction of sunspot numbers from 1 to 12 months ahead based on the standard curve method \citep{Waldmeier1968}, the combined method \citep{Denkmayr1997, Hanslmeier1999}, the McNish–Lincoln method \citep{McNish1949}, and their further improvements with an adaptive Kalman filter \citep{Podladchikova2012}. The modification of the McNish–Lincoln method suggested by \cite{Holland1984,Niehuss1996} is used for the operational medium-term prediction of solar radio flux at F10.7~cm within SOLMAG, which is the model for predicting solar and geomagnetic activity employed by ESA’s Space Debris Office \citep{Mugellesi1993,Virgili2014}. \citet{Petrova2021} developed a new technique (RESONANCE) for the prediction of the solar F10.7~cm and F30 cm radio fluxes with lead times from 1 to 24 months by combining the McNish–Lincoln method and an adaptive Kalman filter. \citet{Lantos2000} showed that the slope at the inflection point
of the cycle ascending profile, which is closely related to the curvature of the sunspot series, serves as a precursor of the subsequent cycle amplitude. \citet{CameronSchlussler2008} presented that there is a steady relation between the average growth rate of solar activity in the ascending phase of a cycle and the subsequent cycle amplitude. In combination with the Waldmeier effect, i.e. the inverse correlation between cycle rise time and its amplitude \citep{Waldmeier1968}, these findings are indicative of a faster-than-linear increase of the growth rate with cycle amplitude \citep{Karak2011}.

In this study, we introduced a new precursor, namely the maximal growth rate of sunspot activity in the ascending phase of a cycle, and analysed its capacity to predict the amplitude of a subsequent cycle in comparison with the average growth rate indicator. We also investigated whether the suggested precursor provides benefits for the prediction of the solar cycle amplitude when using the sunspot indices (sunspot numbers, sunspot areas) derived separately for the two hemispheres compared to the total sunspot indices describing the entire solar disc. Previous studies demonstrated an asymmetry in the solar cycle evolution and the solar magnetic field in the northern and southern hemispheres \citep{Maunder1904,Spoerer1889,Waldmeier1971,Newton1955, Carbonell1993,Temmer2001,Temmer2002,Berdyugina2003,Durrant2003,Joshi2004,Knaack2005,Norton2010,McIntosh2013,Deng2016,Temmer2006,Zolotova2010,Svalgaard2013,Norton2014,Chowdhury2019,Roy2020}, which is in agreement with dynamo theories \citep{Sokoloff1994,Ossendrijver1996,Charbonneau2005,Norton2014,Schuessler2018}. However, the existing hemispheric sunspot number data series from 1945 onwards \citep{Temmer2006} were too short to be used for predictions of the solar cycle amplitude. This shortcoming was recently overcome by the newly created hemispheric sunspot number catalogue, which covers an extended time range from 1874 onwards \citep{Veronig2021Hemispheric}. This data set is the basis of our analysis.
In addition, we also checked if our findings are confirmed with the hemispheric sunspot areas.

\section{Data preparation} \label{sec_data}
In this study, we used four data sets of monthly solar activity indices: total sunspot numbers, hemispheric sunspot numbers, total sunspot areas, and hemispheric sunspot areas. Total sunspot numbers are provided by the WDC-SILSO\footnote {http://www.sidc.be/silso/} at the Royal Observatory of Belgium (ROB). Hemispheric sunspot numbers are available in the newly created catalogue for the time range 1874--2020 \citep{Veronig2021Hemispheric}\footnote{\url{http://cdsarc.u-strasbg.fr/viz-bin/cat/J/A+A/652/A56}}, which is also distributed by WDC SILSO\footnote{\url{http://www.sidc.be/silso/extheminum}}. The catalogue of \citet{Veronig2021Hemispheric} provides two slightly different sets of hemispheric sunspot numbers, due to different data sources that are used in the merging of the time series (for periods with overlapping data). The first set consists of hemispheric data merged from three different sources: reconstructed from sunspot areas over 1874--1945, derived and re-calibrated from \citet{Temmer2006} over 1945-1991, and compiled from WDC-SILSO over 1992-2020. Throughout the text, we refer to them as `merged hemispheric sunspot numbers'. The second set consists of hemispheric data reconstructed purely from sunspot areas over 1874--2016 (shown in Figure~\ref{fig9}), and we refer to them as `full-proxy hemispheric sunspot numbers'. Total and hemispheric sunspot areas are provided by the Royal Observatory, Greenwich - USAF/SOON and NOAA\footnote{\url{https://solarscience.msfc.nasa.gov/greenwch.shtml}}.

\begin{figure*}  
        \centering
        \includegraphics[width=12cm]{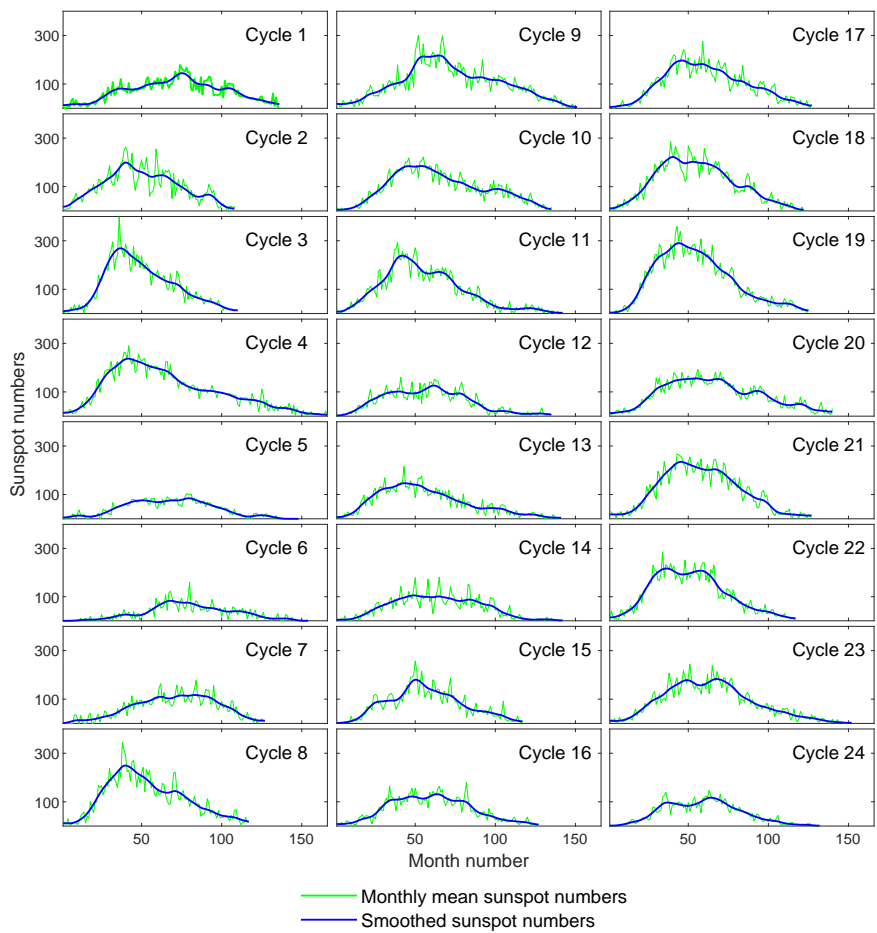}
        \caption{Smoothed monthly mean sunspot numbers (blue) for cycles 1–24 using a 37-month optimised running mean. The green line shows the monthly mean sunspot number.} 
        \label{fig1}
\end{figure*}

In this study, we calculated the growth rate of sunspot activity in the ascending phase of a cycle in terms of sunspot number first differences (SNFD) or sunspot area first differences (SAFD). For reliable estimations of SNFD (or SAFD), which are very sensitive to noise in the data, we propose using the optimised running mean filter, which is advantageous in the study of short-term variations of sunspot activity with respect to the traditional boxcar \citep{Hathaway1999} averaging of the monthly mean data, as demonstrated in \citet{Podladchikova2017}. To this aim, we derived the smoothed values by minimising the following functional: 
\begin{equation}\label{eq_functional_J}
J = \beta \sum_{i=1}^{n}(S_{i}^{m}-{S}_{i})^2+\sum_{i=1}^{n-2}(S_{i+2}-2S_{i+1}+S_{i})^2
.\end{equation}
Here, $S_{i}^{m}$ is the monthly mean index of solar activity (total or hemispheric sunspot numbers, total or hemispheric sunspot areas), $S_{i}$ is the smoothed index in the month, $i$, and $\beta$ is a smoothing constant, which regulates how closely the smoothed curve fits the data. For our study, we use 
$\beta = 0.01$, which provides an effective noise filtration and reliable estimation of intrinsic variations of a solar cycle as shown in \citet{Podladchikova2017}. The proposed optimised running mean filter is based on finding a balance between the fidelity to the data, as assessed by minimising the sum of the squared deviations between the fit and the data (first term in Equation~(\ref{eq_functional_J})), and the smoothness of an approximating curve, which is assessed by minimising the sum of squared second derivatives of the fit curve (second term in Equation~(\ref{eq_functional_J})). The smoothed values $S_{i}$, which minimise the functional $J$ are derived by solving a system of $n$ normal equations with $n$ unknowns, where $n$ is the number of available monthly mean data. The technique has been further adapted to be used as the optimised running mean filter (see \citet{Podladchikova2017}).

\begin{figure*}  
    \centering
    \includegraphics[width=12cm]{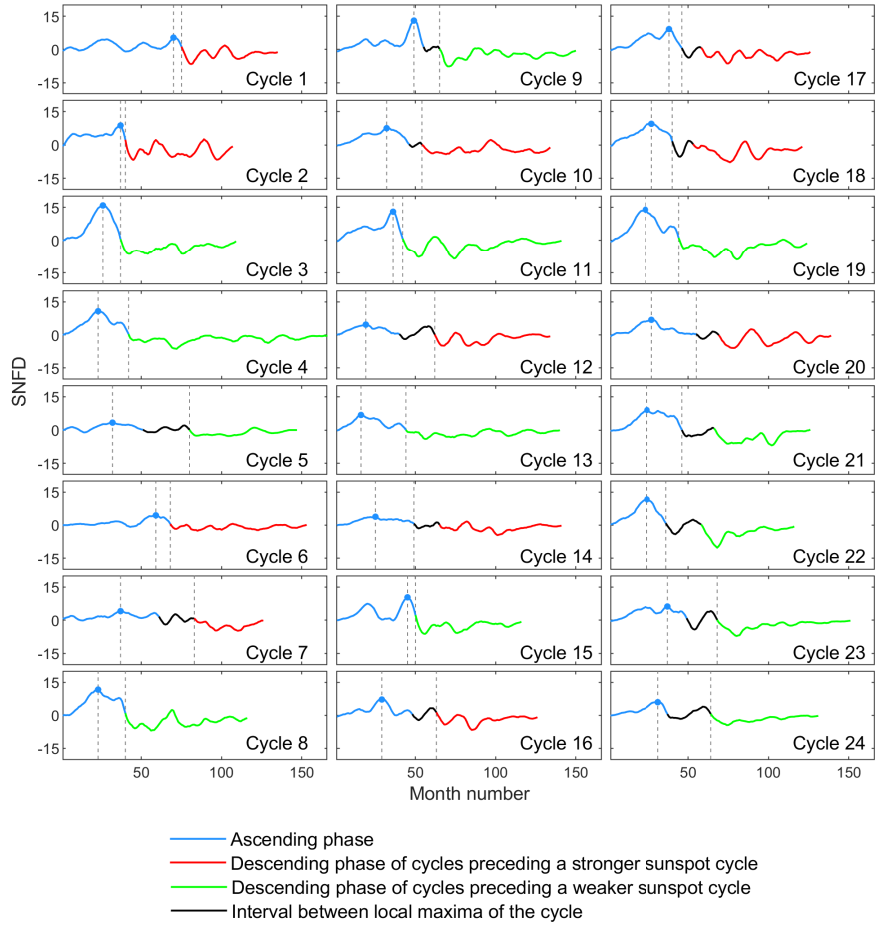}
        \caption{Sunspot number first differences (SNFD) for cycles 1--24 using the 37-month optimised running mean smoothing. The ascending phase of each cycle to its first peak is shown by the blue lines. The SNFD peak representing the maximal growth rate in the ascending phase is marked by the blue dot. The descending phases of cycles that precede a stronger (weaker) cycle are indicated by red (green) lines. If a cycle shows two or more peaks, then the interval between the first and the latest peak is marked by a magenta line. The same plots, but for a 13-month smoothing window, 
        are shown in Appendix~\ref{Appendix_A} in Figure~\ref{fig11}.}
        \label{fig2}
\end{figure*}

To study the influence of window size of the optimised running mean on the prediction accuracy of cycle amplitudes, we considered the following range of smoothing windows: 7, 9, 11, 13, 15, 21, 23, 27, 33, 37, 41, and 45 months. The larger the window, the smoother the resulting approximation curve, and vice versa. Figure~\ref{fig1} shows the monthly mean (green) and smoothed (blue) total sunspot numbers with a 37-month optimised running mean for cycles 1--24. In the same way, we also processed the hemispheric sunspot numbers and the total and hemispheric sunspot areas. 

\section{Growth rate of total sunspot numbers as a precursor of the sunspot cycle amplitude} \label{sec_precursor}
We analysed the relationship between the growth rate indicators in the ascending phase of a sunspot cycle and the subsequent cycle amplitude. One indicator, used by \citet{CameronSchlussler2008}, is the 'average growth rate' determined in the empirically chosen interval of 30--50 for the total sunspot number in the ascending phase of a cycle. In this study, we had to take into account the recent re-calibration of sunspot numbers \citep{Clette2016b, Clette2016}, where the scale of the index is increased, on average, by a factor of 1.4. Therefore, we used the equivalent sunspot number interval 42--100 to derive the mean growth rate. In addition, we introduced a new indicator, namely the 'maximal growth rate' estimated as the SNFD peak in the ascending phase. SNFD are defined as the first differences $S_{i}-S_{i-1}$ of the sunspot numbers smoothed with the optimised running mean filter described in Section~\ref{sec_data}.
%

Figure~\ref{fig2} shows the SNFD for cycles 1--24 using the monthly mean data smoothed with a 37-month window. Figure~\ref{fig11} in Appendix~\ref{Appendix_A} shows the same, but for a 13-month smoothing window. The ascending phase of each cycle to its first peak is shown as a blue line, and the corresponding SNFD peak is marked by blue dots. At the beginning of the descending phase of a solar cycle, the SNFD changes sign from positive to negative. Red lines indicate the descending phase of cycles that precede stronger cycles, whereas green lines indicate the descending phase of cycles that precede a weaker cycle. If a cycle is characterised by two or more peaks, the interval between the first and the latest peak is marked by a black line. Black dashed vertical lines indicate the time of the SNFD peak and cycle peak, respectively.
\begin{figure}  
   \centering
   \includegraphics[width=1\columnwidth]{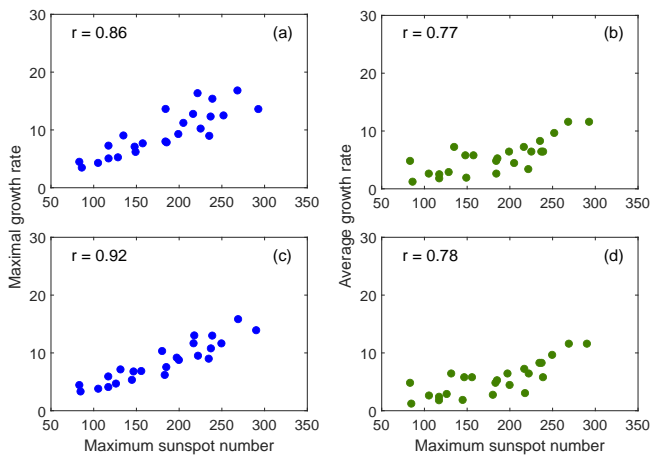} 
        \caption{Relation between the growth rate indicators and the amplitudes of solar cycles 1--24. Left panels~(a)~and~(c): maximal growth rate (SNFD peaks). Right panels~(b)~and~(d): average growth rate. Top panels~(a)~and~(b): 13-month optimised running-mean. Bottom panels~(c)~and~(d): 37-month optimised running mean. The numbers indicate the solar cycle number.}
        \label{fig3}
\end{figure}

Figure~\ref{fig3} shows the dependence of the growth rate indicators on the solar cycle amplitude. The left panels show the results for the maximal growth rate (SNFD peaks), and the right panels show results for the average growth rate. Top (bottom) panels refer to 13-month (37-month) smoothing windows. The correlation coefficients $r$ between the growth rate indicators and the solar cycle amplitudes show a higher performance for the maximal growth rates ($r=0.92$) than for the mean growth rates ($r=0.86$), whereas they are similar for the 13-month smoothing ($r=0.77$ and $r=0.78$, respectively).

\begin{figure*}  
   \centering
    \includegraphics[width=1.7\columnwidth]{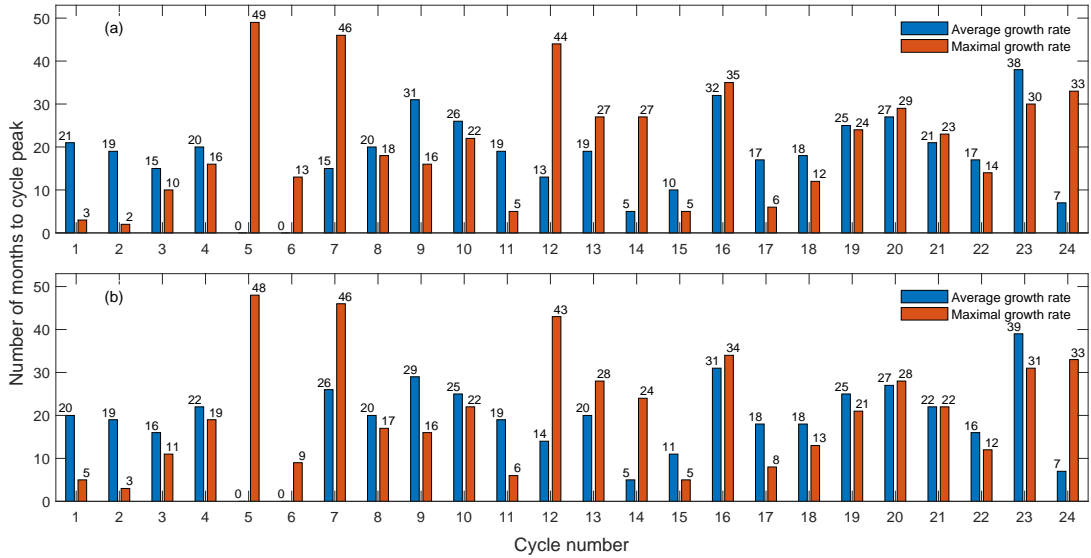}
        \caption{Time difference (in months)     between time of estimating the growth rate indicators and cycle amplitudes. (a) 13-month and (b) 37-month smoothings. Red bars represent the maximal growth rate (SNFD peak), and blue bars represent the average growth rate.}
        \label{fig4}
  \centering
\end{figure*}

Figure~\ref{fig4} shows how early in the ascending phase of a cycle we can estimate the growth rate indicators. The x-axis gives the cycle number, while the y-axis indicates the number of months between the time of estimating the growth rate indicators and the time where the solar cycle reaches its amplitude. This is a relevant parameter when the growth indicators are used for solar cycle predictions. As can be seen from Figure~\ref{fig4}, for 14 solar cycles (1, 2, 3, 4, 8, 9, 10, 11, 15, 17, 18, 19, 22, and 23) the average growth rate indicator (estimated when the smoothed sunspot number reached a value of 100) occurs earlier than the maximal growth indicator, whereas in the other ten cycles (5, 6, 7, 12, 13, 14, 16, 20, 21, and 24) the maximal growth indicator occurs earlier. The first group of cycles, for which the average growth rate occurs earlier than the maximal growth rate, is formed of strong cycles, with a straight or concave rising profile. The second group, for which the SNFD peak precedes the average growth rate, is formed of weaker cycles, with a more progressive transition between the rising part and a rather broad maximum, giving a convex rising profile. The lead times vary from 0 to 39 months for the average growth rate and from 2 to 49 months for the maximal growth rate, respectively. The mean lead time is the same for both indicators: 18 (21) months for the 13-month (37-month) smoothed data. Finally, we note that for cycles 5 and 6 it would not be possible to use the average growth rate indicator, since their amplitudes are smaller than 100.
\begin{figure}
    \centering
    \includegraphics[width=1\columnwidth]{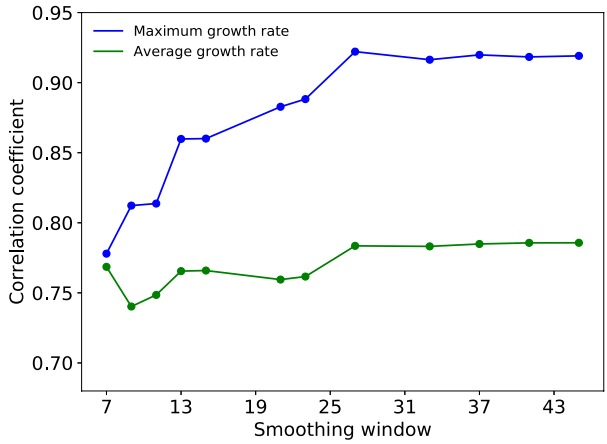}
        \caption{Correlation coefficient between growth rate indicators and cycle amplitudes for different smoothing windows        from 7 to 45 months. The blue line shows the relation for the maximal growth rate (SNFD peak), and the red line shows the average growth rate.}
        \label{fig5}
\end{figure}

In Figure~\ref{fig5}, we can see the dependence of the relation between the growth rate indicators and the cycle amplitudes on the smoothing level applied to the data. The plot shows the correlation coefficient between the two quantities for different smoothing windows from 7 to 45 months. The blue line shows the results for the maximal growth rate, and the red lines show results for the average growth rate. Figure~\ref{fig5} shows two important results. Firstly, the higher the degree of smoothing, the larger the correlation coefficient; starting from a smoothing window of 27 months, the correlation coefficient saturates, indicating that stronger smoothing no longer improves the relation. Secondly, the correlation coefficient is always higher for the maximal than for the average growth rate indicator. For the maximal growth rate it increases from $r=0.78$ to $r=0.92$ for smoothing windows from 7 to 27 months, whereas for the average growth rate it rises only slightly from 0.74 to 0.78. While the values of the average growth rate are less variable for different smoothing windows, the SNFD peaks for higher degrees of smoothing lead to an increase of the correlation coefficient due to the effective filtration of SNFD fluctuations.  

\section{Predicting the solar cycle amplitude with total sunspot numbers} \label{sec_prediction_tsn}

In this section, we analyse how the growth rate indicators can be used for the predictions of the solar cycle amplitudes. To this aim, we created the following third-order linear regression:
\begin{equation}\label{eq_regression}
S^{p} = \beta_{0} + \beta_{1}I + \beta_{2}I^2 + \beta_{3}I^3 
.\end{equation}
Here, $S^{p}$ denotes a value of the cycle amplitude, $I$ is a value of the growth rate indicator (maximal or average). The vector of regression coefficients 
$\left|\beta_{0},\ \beta_{1},\ \beta_{2}, \ \beta_{3} \right|^{T}$ is determined from the least-squares method for both growth rate indicators and all the considered smoothing windows. We note that despite the third order implying a non-linear relationship between $S^{p}$ and $I$, the regression defined by Equation~(\ref{eq_regression}) is still considered to be linear, as it is linear with respect to unknown vector of regression coefficients estimated from the data.
\begin{figure}
    \centering
    \includegraphics[width=1\columnwidth]{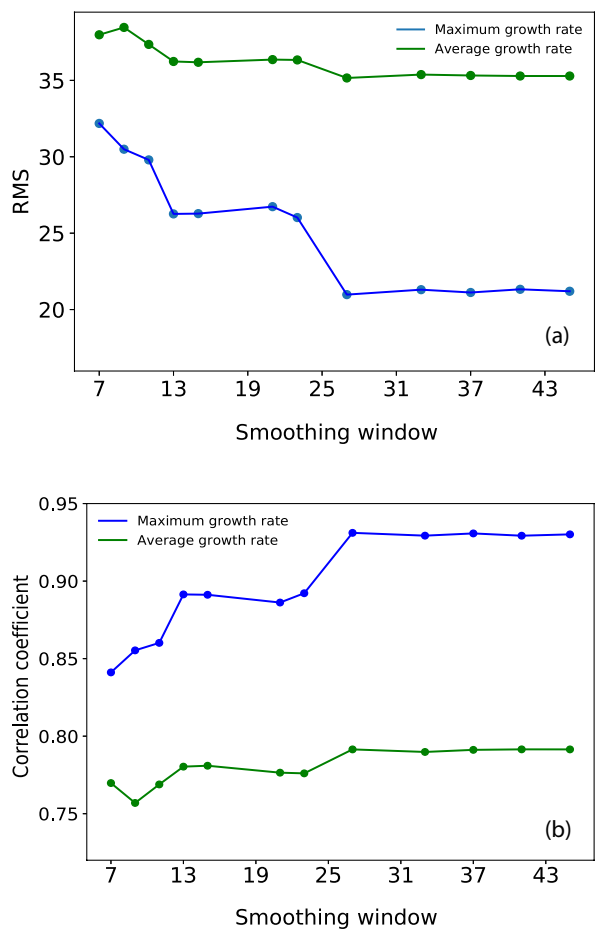}
        \caption{Prediction performance of amplitudes of cycles 1--24 for different smoothing windows from 7 to 45 months.}
        \label{fig6}
\end{figure}

Figure~\ref{fig6} shows the prediction performance for cycles 1--24. Panel~(a) gives the rms errors of the predictions, and panel~(b) indicates the correlation coefficient between the true and predicted cycle amplitudes with the maximal growth rate (blue) and average growth rate (green). As can be seen from Figure~\ref{fig6}, the larger the smoothing window, the smaller the rms error of predictions, and the higher the correlation between the true and predicted cycle amplitudes. The rms error of the predictions using the maximal growth rate indicator decreases from 32 to 21. On average, it is smaller (by 31\%) than that for the average growth rate, which only reduces from 39 to 35. For the maximal growth rate indicator, the correlation coefficient between the true and predicted cycle amplitudes increases from $r =0.8$4 to $r=0.93$, and shows, on average, a performance superior by 15\% to the average growth rate, where $r$ only rises from 0.76 to 0.79. The highest accuracy is already achieved for the smoothing window of 27 months. This is important for the prediction lead time, taking into account that the values of the growth rate indicators are estimated with a delay of half of the chosen smoothing window. 

The increase of the smoothing window from 7 to 27 months leads to a more effective filtration of fluctuations in the SFND, and thus to an increase of the prediction accuracy. From one side, smaller sizes of smoothing windows allow us to estimate the growth rate indicators earlier than the larger one. However, as is shown in Figure~\ref{fig6}, it slightly decreases the accuracy of the predictions. From another side, as discussed with respect to  Figure~\ref{fig4}, the SNFD peak is, on average, determined earlier for larger smoothing windows. 
\begin{table*}
        \centering
        \caption{Prediction lead times.}
    \begin{tabular}{l c c c c c c c c c c c c c} 
                \hline\hline 
                & \textbf{1} & \textbf{2} & \textbf{3} & \textbf{4} & \textbf{5} & \textbf{6} & \textbf{7} & \textbf{8} & \textbf{9} & \textbf{10} & \textbf{11} & \textbf{12}  \\ 
                \hline 
                60\% & 53.9 & 33.7 & 20.6 & 27.1 & 57.4 & 25.5 & 50.8 & 25.2 & 21.1 & 38.1 & 25.8 & 52.2 \\ 
                70\% & 52.7 & 9 & 18.9 & 24.8 & 56.3 & 17 & 49.7 & 24.1 & 20.7 & 34.2 & 11 & 50.6\\
                80\% & 45.2 & 6.6 & 17.7 & 23.3 & 55.1 & 16.5 & 48.7 & 22.6 & 20.1 & 25 & 9.9 & 48.1\\
                90\% & 8.4 & 5.2 & 16.1 & 21.4 & 51.4 & 16 & - & 19.6 & 19.3 & 23.5 & 9 & 44.8\\ 
                \hline\hline
                & \textbf{13} & \textbf{14} & \textbf{15} & \textbf{16} & \textbf{17} & \textbf{18} & \textbf{19} & \textbf{20} & \textbf{21} & \textbf{22} & \textbf{23} & \textbf{24}  \\ 
                \hline
                60\% & 34 & 34.4 & 35.4 & 42.3 & 27.8 & 22.7 & 29.5 & 38.7 & 28.1 & 21.3 & 52.6 & 43.6\\ 
                70\% & 33.1 & 32.7 & 34.3 & 40 & 26.1 & 20.5 & 28.2 & 37.6 & 26.6 & 19.9 & 47.9 & 41.8\\
                80\% & 32.2 & 31.6 & 33.2 & 39.5 & 22.8 & 18.2 & 26.5 & 36.2 & 24.2 & 18.9 & 35.3 & 40.6\\
                90\% & 31.2 & 26.7 & 31.8 & 38.9 & 13.2 & 16.2 & - & 35 & - & 17.6 & - & 39.7\\ [1ex] 
                \hline 
    \end{tabular}
        \label{table1} 
        \tablefoot{Lead time (in months) when the prediction using increasing SNFD in the ascending phase (for 37-month smoothing window) reached 60, 70, 80, and 90\% from the actual solar cycle amplitude for cycles 1--24. Cycle numbers are indicated in bold.}
\end{table*}

We can further adapt the proposed prediction technique to be used in real-time and predict the cycle amplitude continuously over the development of the ascending phase of a solar cycle. In this case, we did not wait until the maximum of the growth rate is reached, but we updated the prediction when the latest value of the growth rate is larger than the previous one. Table~\ref{table1} shows the forecast lead time in months when the prediction with the increasing SNFD in the cycle ascending phase, estimated on the basis of a 37-month optimised running mean, reached 60, 70, 80, and 90\% from the actual solar cycle amplitude for cycles 1--24. As is shown in Table~\ref{table1}, the average lead time for 60\% of a cycle amplitude is 36.4 months and varies from 20.6 to 57.4 months. For 70\% of a cycle amplitude, it is 32.5 months, for 80\% it is 28.6 months, and for 90\% it is 24.5 months.

Using the SNFD peak obtained from the 13-month smoothed sunspot numbers of the rising phase of ongoing cycle 25, which occurred 23 months after its start (Figure~\ref{fig11}), we predict that the lower estimate of the amplitude of solar cycle 25 will be 110$\pm$26, which is in line with other predictions that lie in the range from around 90 to 139  \citep{Kitiashvili2016, Podladchikova2017, Singh2017, Bhowmik2018, Labonville2019, Miao2020, Burud2021, Kumar2021, Brajsa2022}.  However, it is in contrast with the study by \citet{McIntosh2020},  who predicted a very large amplitude of 233 for cycle 25 based on termination events that mark the end of a solar cycle. 

\section{Predicting the solar cycle amplitude with hemispheric sunspot data} \label{sec_prediction_hs}
In this section, we investigate whether sunspot activity indices derived separately for the two hemispheres can benefit the predictions of solar cycle amplitudes in comparison with the sunspot indices describing the entire solar disc. We repeated the same steps as performed for total sunspot numbers with regard to hemispheric sunspot numbers and total and hemispheric sunspot areas.
\begin{figure*} 
    \centering
    \includegraphics[width=1.5\columnwidth]{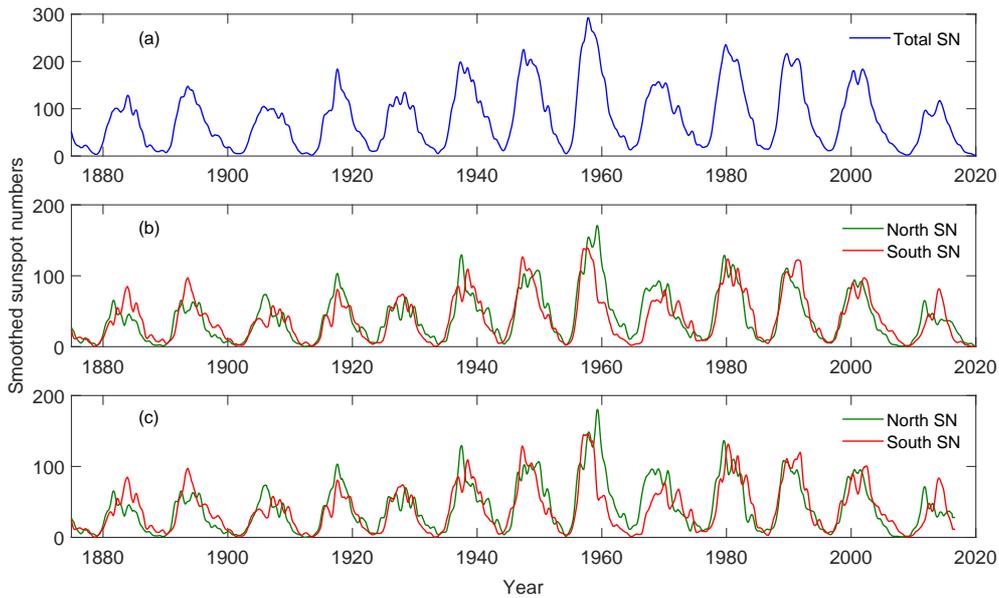}
        \caption{Total and hemispheric sunspot numbers (37-month smoothed). (a) Total, (b) merged, and (c) full-proxy hemispheric sunspot numbers (green:  north; red: south; blue: total).}
        \label{fig7}
\end{figure*}

Figure~\ref{fig7} shows the 37-month smoothed  sunspot numbers over the 1874--2020 time range, covering solar cycles from 12 to 24. Panel (a) shows the total sunspot numbers, (b) the merged hemispheric sunspot numbers, and (c) the full-proxy hemispheric sunspot numbers. For the predictions of the solar cycle amplitudes using the hemispheric data, we again used the third-order linear regression defined by Equation~(\ref{eq_regression}), where $S^{p}$ denotes the value of the cycle amplitude based on total sunspot numbers, but $I$ is the sum of SNFD peaks determined separately for the northern and southern hemispheric sunspot numbers.
\begin{figure*}
    \centering
    \includegraphics[width=1.9\columnwidth]{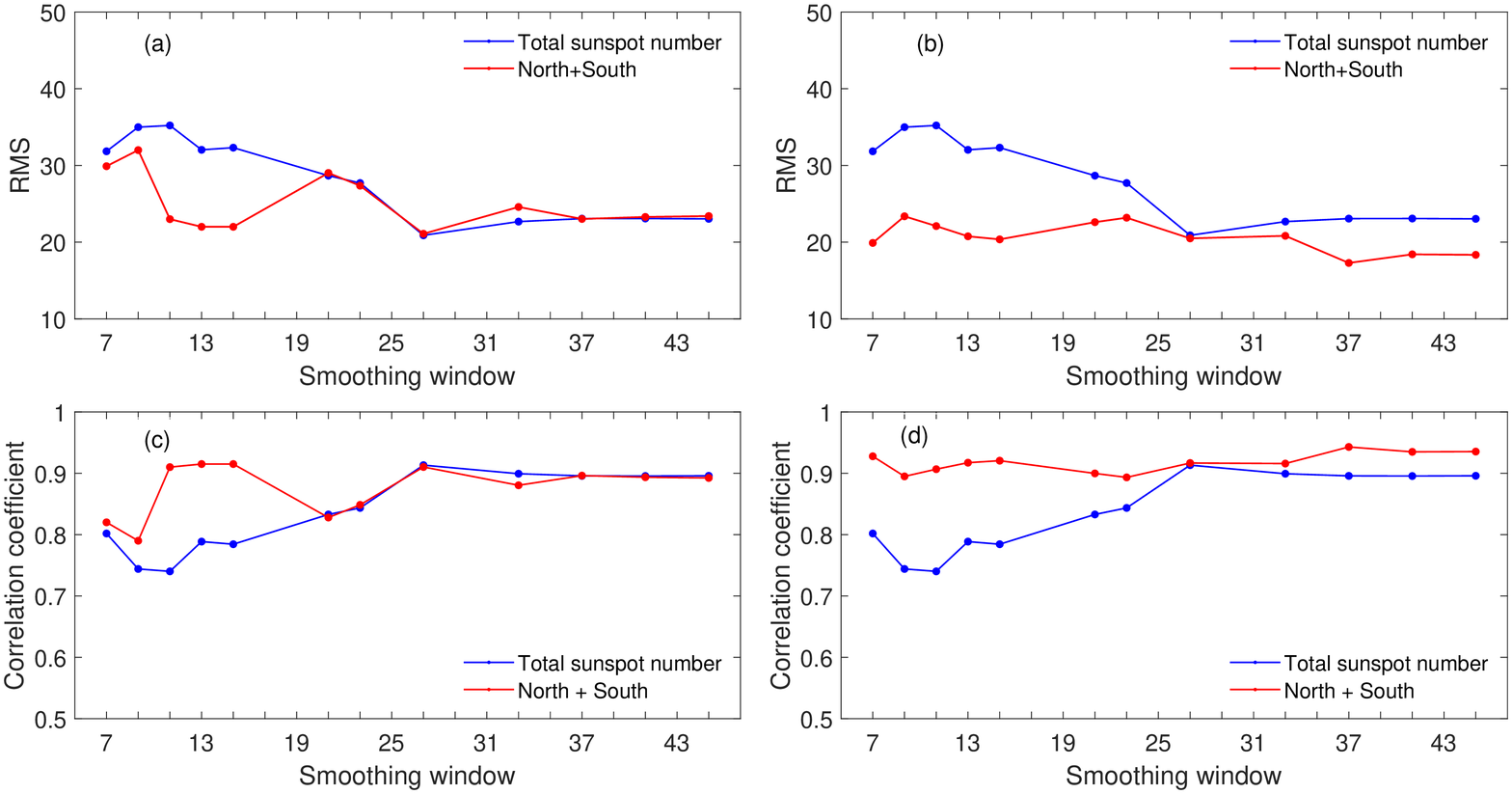}
        \caption{Amplitude prediction performance of solar cycles {12--24} with the total (blue) and hemispheric (red) sunspot numbers for different smoothing windows (from 7 to 45 months). Top panels: Rms errors of  prediction. Bottom panels: Correlation coefficient between true and predicted cycle amplitudes. Left (right) panels: Prediction using the merged (full-proxy) hemispheric sunspot numbers. }
        \label{fig8}
\end{figure*}

Figure~\ref{fig8} shows the prediction performance for cycles 12--24 based on the hemispheric sunspot numbers (red). For comparison, we also indicate the prediction accuracy using the total sunspot numbers (blue). We note that there are now fewer cycles entering the calculation. This is because the hemispheric data only exist over 13 available solar cycles, which, on average, reduces the accuracy by 9\% and 8\% in terms of rms and r, respectively, compared the same results using all the cycles 1-24 (Figure~\ref{fig6}, blue lines). Top panels~(a)~and~(b) show the rms errors of predictions; bottom panels~(c)~and~(d) show the correlation coefficient between the true and predicted cycle amplitudes. Left panels (a) and (c) represent the results for the merged hemispheric sunspot numbers; right panels (b) and (d) represent the results for the full-proxy hemispheric sunspot numbers.

The left panels in Figure~\ref{fig8} show that for smaller smoothing windows (from 7 to 15 months) the predictions of cycle amplitudes using the sum of the north and south SNFD peaks of the merged hemispheric sunspot numbers show a superior performance in terms or rms compared to the total sunspot numbers, whereas for larger smoothing windows the performances are similar. The rms error of the predictions with the merged hemispheric sunspot numbers decreases from 32 to 21 for smoothing windows of 7 to 15 months, and it is, on average, smaller than that for the total sunspot numbers by 9\%, which decreases from 35 to 21. The correlation coefficient between the true and predicted cycle amplitudes with the merged hemispheric sunspot numbers increases from $r=0.79$ to $r=0.90$ and is, on average, 15\% higher than that for the total sunspot numbers where $r$ increases from 0.74 to 0.91 over the same smoothing range. 

However, as can be seen in the right panels of Figure~\ref{fig8}, there is a stronger outcome with the full-proxy hemispheric data sets, which exhibit an overall superior performance for all the smoothing windows compared to the total sunspot numbers (rms is smaller by 27\%, and $r$ is higher by 11\%, on average). Interestingly, both the rms error and the correlation coefficient show a very stable and high-performance behaviour for the different smoothing windows. For all window sizes, the correlation coefficient is $r\ge 0.9$ and rms $<23$. This more robust and higher performance of the full-proxy hemispheric sunspot numbers as compared to the merged ones (in particular for small smoothing windows) is probably related to the better homogeneity of the hemispheric sunspot time series, which is only based on one underlying time series (hemispheric areas) in contrast to the one merged from different data sets. However, we note that the merged series has the advantage that it is continuously updated with the new data and consistent with the SILSO hemispheric sunspot numbers, whereas the hemispheric sunspot areas from Greenwich are no longer available. With the available merged 13-month smoothed hemispheric sunspot data, we predict that the lower estimate of the amplitude of cycle 25 will be 93$\pm$21.

\begin{figure*}
    \centering
    \includegraphics[width=1.5\columnwidth]{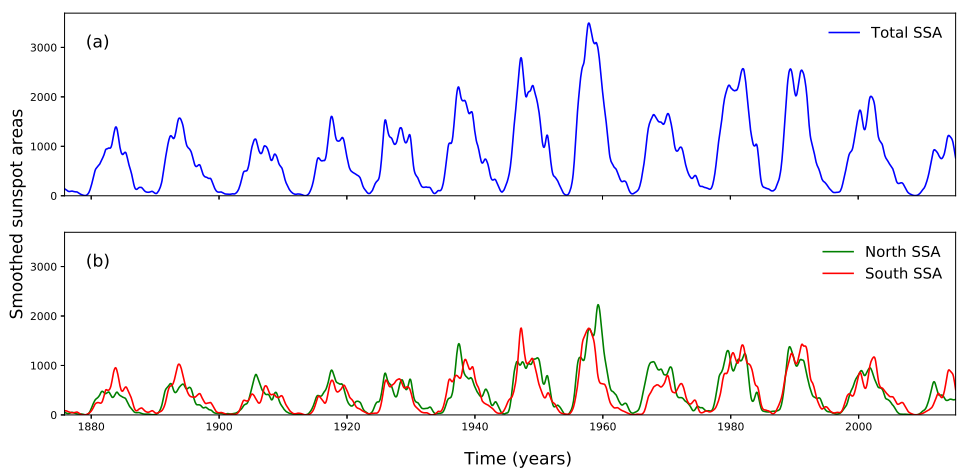}
        \caption{ Total and hemispheric sunspot areas (37-month smoothed) in units of millionths of a solar hemisphere.  (a) Total, (b) merged, and (c) full-proxy hemispheric sunspot numbers (green:  north; red: south; blue: total).}
        \label{fig9}
\end{figure*}

Figure~\ref{fig9} shows the evolution of the 37-month smoothed total and hemispheric sunspot areas over the 1874--2016 time range. For the predictions of solar cycle amplitudes, we again used Equation~(\ref{eq_regression}), but now $S^{p}$ denotes the value of the cycle amplitude in terms of the total sunspot areas and $I$ is the sum of the SNFA growth rate peaks determined separately for the northern and southern hemispheric sunspot areas.
\begin{figure}
    \centering
    \includegraphics[width=0.88\columnwidth]{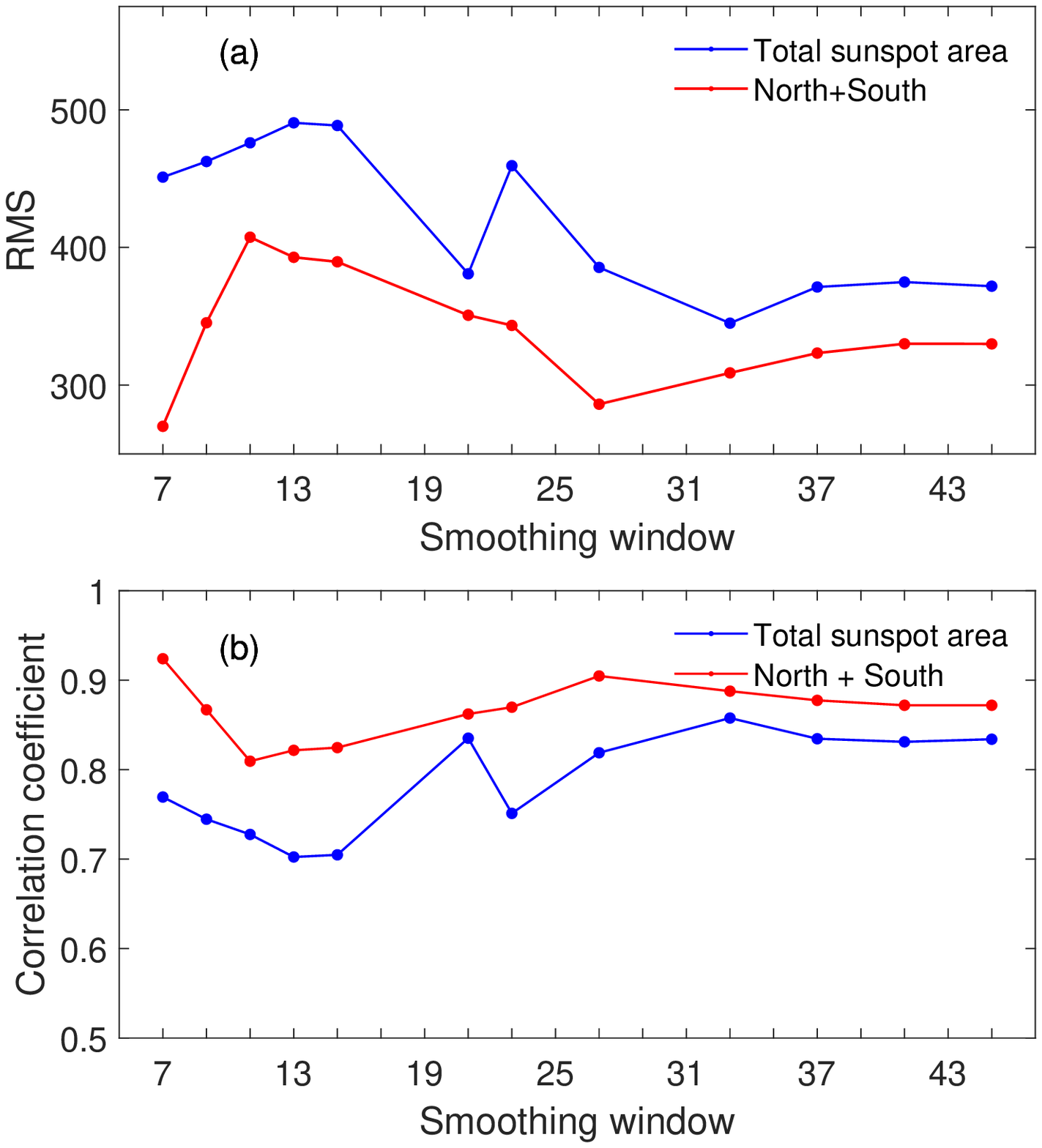}
        \caption{Prediction performance for cycles 12-24 with the total (blue) and hemispheric (red) sunspot areas. (a) rms errors of predictions, (b) correlation coefficient between the actual cycle amplitudes and predictions. X-axis: Width of smoothing window from 7 to 45 months.}
        \label{fig10}
\end{figure}

Figure~\ref{fig10} shows the prediction performance for cycles 12--24 using the total (blue) and the hemispheric sunspot areas (red). Panel~(a) shows the rms errors of the predictions and panel (b) the correlation coefficient between the true and predicted cycle amplitudes. As can be seen from Figure~\ref{fig10}, the prediction accuracy for the sunspot areas remains lower than that one for sunspot numbers for all smoothing windows (Figure~\ref{fig10}). At the same time, the predictions of cycle amplitudes using the sum of the north and south SNFA peaks of the hemispheric sunspot areas in the ascending phase of a cycle again show superior performance to those using the total sunspot areas, for all smoothing windows. The decrease in prediction accuracy with an increase of the smoothing window from 7 to 11 months may be random due to the small sample. The rms error of the predictions with the hemispheric sunspot areas decreases from 407 to 270 units and, on average, is smaller by 19\% than that for the total sunspot areas, which decreases from 491 to 345 units. The correlation coefficient between the true and predicted cycle amplitudes with the hemispheric sunspot areas varies from $r=0.81$ to $r=0.92$, which is, on average, 11\% higher than that for the total sunspot areas, where $r$ varies from 0.70 to 0.86.

\section{Summary and conclusion}\label{sec_conclusion}
The findings of this study can be divided into three main groups. First, we showed that the maximal growth rate of sunspot activity in the ascending phase of a solar cycle is a better precursor of a subsequent solar cycle amplitude than the average growth rate. Second, we developed and tested a prediction technique based on using the maximal growth rate as a precursor. Finally, we demonstrated that the hemispheric sunspot indices derived separately for the two hemispheres provide advantages in predicting the solar cycle amplitudes compared to the sunspot indices describing the entire solar disc. 

We demonstrated a steady relation between the maximal growth rate of activity in the ascending phase of the cycle and the subsequent cycle amplitude for four data sets of global activity indices: total sunspot numbers describing the full Sun, hemispheric sunspot numbers from the new catalogue from 1874 onwards \citep{Veronig2021Hemispheric}, total sunspot areas, and hemispheric sunspot areas. The growth rate of sunspot indices was estimated as first differences of smoothed sunspot numbers (SNSD) and smoothed sunspot areas (SAFD). To process the noisy record in sunspot data for a robust estimation of the SNFD and SAFD series and the related growth rate indicators, we applied an optimised running mean filter \citep{Podladchikova2017}. The application to solar cycles 1--24 showed that the maximal growth rate defined by SNFD peaks outperforms the average growth rate (on average by 14\%) in terms of the correlation coefficient between the growth rate indicators and the cycle amplitude. 

We used the SNFD peaks for the predictions of the solar cycle amplitudes based on a third-order linear regression technique. Testing this prediction method for cycles 1-24 showed that the rms error of predictions is in the range from 32 to 21, and the correlation coefficient between the true and predicted cycle amplitudes varies from $r=0.84$ to 0.93 depending on the size of the smoothing window. The lead time of the predictions varies from 2 to 49 months, with a mean value of 21 months. The prediction technique was further adapted for real-time application to forecast the cycle amplitudes over the ongoing development of the cycle ascending phase. In this case, we continuously update the prediction at each time step where the value of the growth rate is larger than the previous one. With the available 13-month smoothed sunspot data up until October 2021 of the rising phase of ongoing cycle 25, we derive a value of 110$\pm$26 as a lower estimate of its amplitude.

We also showed that the sum of the SNFD peaks determined separately for the northern and southern hemispheric sunspot numbers provides more accurate predictions of the cycle amplitudes than the same prediction technique using the total sunspot numbers. Validating the predictions for cycles 12--24 reveals that the advantages of using hemispheric sunspot numbers reach, on average, 27\% in terms of the rms prediction errors and 11\% in terms of the correlation coefficient between the predicted and true cycle amplitudes. For large smoothing windows, the correlation coefficient reaches values of to $r=0.94$ and rms as low as 17. According to our method, based on the available 13-month smoothed hemispheric sunspot data in the rising phase of ongoing cycle 25, we predict that its amplitude would exceed 93$\pm$21. The benefit of the hemispheric sunspot indices for the predictions of the solar cycle amplitudes is also confirmed when using the hemispheric sunspot areas. The rms error of predictions with the hemispheric sunspot areas for cycles 12-24 is smaller by 19\% on average, and the correlation coefficient between the true and predicted cycle amplitudes is higher by 11\%, on average, than that for the total sunspot areas. 

These findings provide a strong foundation for supporting regular monitoring, recording, and predictions of solar activity based on hemispheric sunspot data. Our findings demonstrate that solar cycle prediction methods can be enhanced by investigating the solar cycle dynamics in terms of the hemispheric sunspot data and by accounting for the different evolution of the two hemispheres over a solar cycle, which in general do not evolve in phase. These two (independent) dynamics of the hemisphere are no longer captured when considering the total sunspot numbers or sunspot areas, which therefore have a lower prediction capacity than the hemispheric indices.

\begin{acknowledgements}
The authors acknowledge the support of the Belgian Solar-Terrestrial Centre of Excellence (STCE) to the World Data Centre SILSO for the production of the sunspot number, and the Royal Observatory, Greenwich - USAF/SOON and NOAA for the total and hemispheric sunspot areas. This research has received financial support from the European Union’s Horizon 2020 research and innovation program under grant agreement No. 824135 (SOLARNET). We thank the referee for valuable comments on this study.
\end{acknowledgements}

\bibliographystyle{aa}
\bibliography{My_References}

\begin{appendix}
\section{Sunspot number first differences based on the 13-month optimised running mean} \label{Appendix_A}
\begin{figure}  
    \includegraphics[width=12cm]{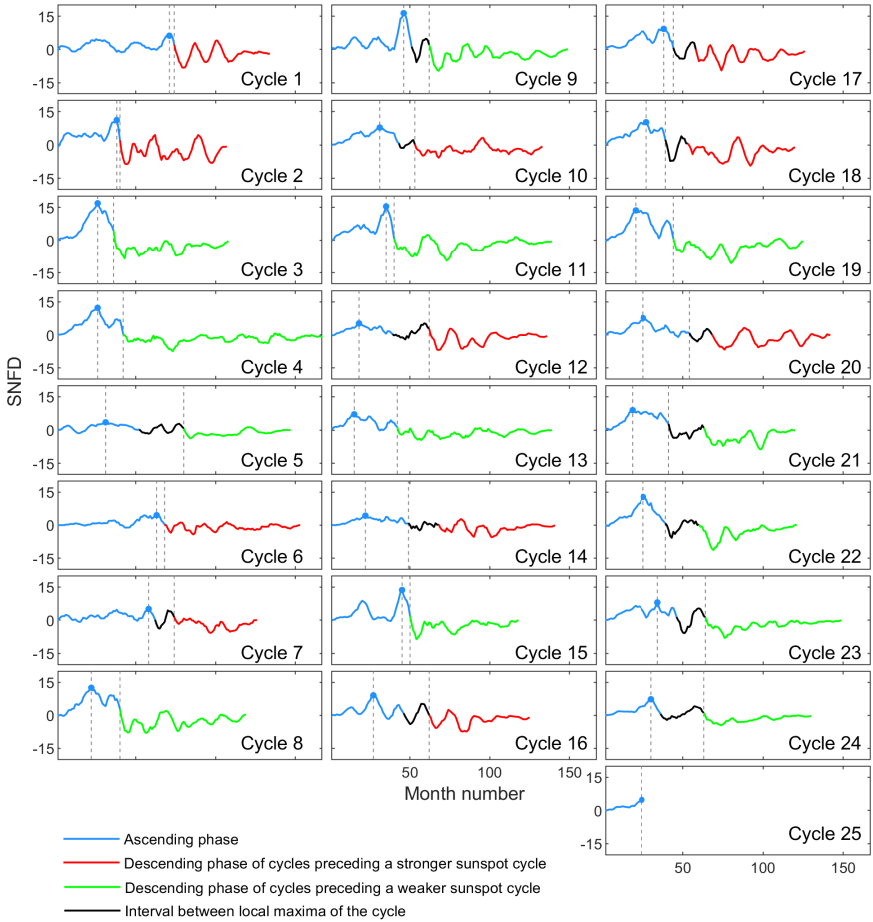}
        \caption{Same as Figure~\ref{fig2}, but for the 13-month optimised running mean.}
        \label{fig11}
\end{figure}
Figure~\ref{fig11} shows the SNFD for cycles 1-24 and the early stage of the rising phase of cycle 25 using the 13-month optimised running mean, which is similar to Figure~\ref{fig2}, where SNFD is represented using the 37-month optimised running mean. As can be seen from Figure~\ref{fig11}, the SNFD with a smaller degree of smoothing (13-month optimised running mean) are characterised by higher scatter than that for the higher level of smoothing (37-month optimised running mean). From one side, it leads to a lower accuracy of cycle amplitudes predictions, but from another side, it may allow us to estimate the growth rate indicators earlier than for the larger smoothing windows. 
\end{appendix}

\end{document}